\documentclass[sigconf,nonacm]{acmart}

\usepackage{hyperref}

\AtBeginDocument{%
  \providecommand\BibTeX{{%
    \normalfont B\kern-0.5em{\scshape i\kern-0.25em b}\kern-0.8em\TeX}}}

\begin{document}

\title[SchenQL: A QL for bib. data with aggregations and domain-specific functions]{SchenQL: A query language for bibliographic data with aggregations and domain-specific functions}

\author{Christin Katharina Kreutz}
\email{ckreutz@acm.org}
\orcid{0000-0002-5075-7699}
\affiliation{%
  \institution{Trier University}
  \city{Trier}
  \country{Germany}
  \postcode{54286}
}

\author{Martin Blum}
\orcid{0000-0003-0005-6162}
 \affiliation{%
  \institution{Trier University}
  \city{Trier}
  \country{Germany}
  \postcode{54286}
}

\author{Ralf Schenkel}
\email{schenkel@uni-trier.de}
\orcid{0000-0001-5379-5191}
 \affiliation{%
  \institution{Trier University}
  \city{Trier}
  \country{Germany}
  \postcode{54286}
}
\setcopyright{none}

\renewcommand{\shortauthors}{Kreutz et al.}

\begin{abstract}
Current search interfaces of digital libraries are not suitable to satisfy complex or convoluted information needs directly, when it comes to cases such as \textit{"Find authors who only recently started working on a topic"}. They might offer possibilities to obtain this information only by requiring vast user interaction.

We present SchenQL, a web interface of a domain-specific query language on bibliographic metadata, which offers information search and exploration by query formulation and navigation in the system. Our system focuses on supporting aggregation of data and providing specialised domain dependent functions while being suitable for domain experts as well as casual users of digital libraries.

\end{abstract}

\ccsdesc[500]{Information systems~Query languages}
\ccsdesc[300]{Information systems~Digital libraries and archives}
\keywords{domain-specific query language, digital libraries, query formulation}

\maketitle

\section{Introduction}

Nowadays, there is a plethora of digital libraries that provide bibliographic metadata such as the ACM DL\footnote{\url{https://dl.acm.org/}}, Bibsonomy~\cite{DBLP:series/xmedia/HothoJBGKSS09}, dblp~\cite{DBLP:journals/pvldb/Ley09}, Google Scholar\footnote{\url{https://scholar.google.com/}}, Semantic Scholar\footnote{\url{https://www.semanticscholar.org/}}, SpringerLink\footnote{\url{https://link.springer.com/}} or ResearchGate\footnote{\url{https://www.researchgate.net}}. These interfaces support search, exploration and sometimes also simple query formulation beyond keyword-based search, but they fail to easily satisfy more complex information needs, e.g. \textit{"Find the papers written by any authors of two different institutions A and B"} or \textit{"Find authors who only recently started working on a topic"}. 

A way to overcome these limitations of search interfaces of digital libraries is usage of specialised tools such as GrapAL~\cite{DBLP:conf/acl/BettsPA19}\footnote{As of 2022, the GrapAL web interface is no longer supported.}, which allows query formulation via the graph-query language Cypher~\cite{cypher}. The big disadvantage with GrapAL is the prerequisite of all users having to be familiar with that all-purpose query language. 

To help users access bibliographic metadata more easily, we extend our domain-specific query language and user interface SchenQL~\cite{schenqljournal}. 
It focuses on aggregation of data and bibliographic functions, such as \texttt{RELATED KEYWORDS TO "DLs"} or \texttt{PUBLICATIONS WITH HIGHEST CORERANK METRIC}.
SchenQL offers simple language components which are easily learned by domain-experts as well as casual users of digital libraries. The queries resemble natural language and the graphical user interface helps in query construction. The two aforementioned information needs could be expressed by the following queries: \texttt{ PUBLICATIONS WRITTEN BY ANY DISTINCT 2 OF [(PERSON WORKS FOR "University of Pisa"), (PERSON WORKS FOR "National Research Council, Italy")]} and \texttt{PERSONS AUTHORED (PUBLICATIONS ABOUT KEYWORD "DLs") AND AUTHORED NO (PUBLICATIONS ABOUT KEYWORD "DLs" WITH YEAR AT MOST 2019)}.



\begin{table*}
\scriptsize
\begin{tabular}{l|p{14cm}|l}
    left & right & description\\ \hline 
    
    \hline
    \textbf{query} & \textbf{C} | \textbf{J} | \textbf{K} | \textbf{PU} | \textbf{PE} | \textbf{I} | \textbf{F} & query start\\\hline
    
    \textbf{C} & \textit{limit}? \textcolor{orange}{CONFERENCE} \textbf{C$_F$}? & conference anchor\\
    \textbf{C$_L$} & \textit{dblpKey}|\textit{acronym} & conference literals\\
    \textbf{C$_F$} & (\textbf{C$_A$}|\textcolor{orange}{WITH DBLPKEY} \textit{dblpKey}|\textcolor{orange}{NAMED $\sim$}? \textit{name}|\textcolor{orange}{ABOUT KEYWORD} (\textcolor{orange}{(}\textbf{K}\textcolor{orange}{)}|\textbf{K$_L$})|\textcolor{orange}{WITH ACRONYM} \textit{acronym}|\textcolor{orange}{WITH YEAR} \textbf{COMP} \textit{year}|\textcolor{orange}{OF} (\textcolor{orange}{(}\textbf{PU}\textcolor{orange}{)}|\textbf{PU$_L$}))|(\textcolor{orange}{WITH} ((\textcolor{orange}{$\sim$}\textit{rank})? (((\textcolor{orange}{HIGHEST}|\textcolor{orange}{LOWEST}) \textbf{METRIC})|(\textcolor{orange}{LONGEST}|\textcolor{orange}{SHORTEST}) (\textcolor{orange}{name}|\textcolor{orange}{acronym}))|\textbf{METRIC} \textbf{COMP} (\textit{h\_index}|\textit{core\_rank})|(\textcolor{orange}{name}|\textcolor{orange}{acronym}) \textcolor{orange}{LENGTH} \textbf{COMP} \textit{number})) (\textbf{OPS} \textbf{C$_F$})? & conference filters\\\hline
    
    \textbf{J} & \textit{limit}? \textcolor{orange}{JOURNAL} \textbf{J$_F$}? & journal anchor\\
    \textbf{J$_L$} & \textit{dblpKey}|\textit{acronym} & journal literals\\
    \textbf{J$_F$} & (\textbf{J$_A$}|\textcolor{orange}{WITH DBLPKEY} \textit{dblpKey}|\textcolor{orange}{NAMED $\sim$}? \textit{name}|\textcolor{orange}{ABOUT KEYWORD} (\textcolor{orange}{(}\textbf{K}\textcolor{orange}{)}|\textbf{K$_L$})|\textcolor{orange}{WITH YEAR} \textbf{COMP} \textit{year}|\textcolor{orange}{WITH ACRONYM} \textit{acronym}|\textcolor{orange}{WITH VOLUME} \textit{volume}|\textcolor{orange}{OF} (\textcolor{orange}{(}\textbf{PU}\textcolor{orange}{)}|\textbf{PU$_L$}))|(\textcolor{orange}{WITH} ((\textcolor{orange}{$\sim$}\textit{rank})? (((\textcolor{orange}{HIGHEST}|\textcolor{orange}{LOWEST}) \textbf{METRIC})|(\textcolor{orange}{LONGEST}|\textcolor{orange}{SHORTEST}) (\textcolor{orange}{name}|\textcolor{orange}{acronym})) |\textbf{METRIC} \textbf{COMP} (\textit{h\_index}|\textit{core\_rank})|(\textcolor{orange}{name}|\textcolor{orange}{acronym}) \textcolor{orange}{LENGTH} \textbf{COMP} \textit{number})) (\textbf{OPS} \textbf{J$_F$})? & journal filters\\\hline
    
    \textbf{K} & (\textit{limit}? \textcolor{orange}{KEYWORD} \textbf{K$_F$}?)|(\textcolor{orange}{RELATED KEYWORDS TO} ((\textcolor{orange}{(}\textbf{K}\textcolor{orange}{)} (\textcolor{orange}{IN} (\textcolor{orange}{(}\textbf{PU}\textcolor{orange}{)}|\textbf{PU$_L$}))?)|\textbf{K$_L$}))|((\textcolor{orange}{$\sim$}\textit{rank})? \textcolor{orange}{MOST FREQUENT KEYWORDS OF} (\textcolor{orange}{(}\textbf{C}\textcolor{orange}{)}|\textcolor{orange}{(}\textbf{J}\textcolor{orange}{)}|\textcolor{orange}{(}\textbf{PU}\textcolor{orange}{)}|\textcolor{orange}{(}\textbf{PE}\textcolor{orange}{)}|\textbf{C$_L$}|\textbf{J$_L$}|\textbf{PU$_L$}|\textbf{PE$_L$}))& keyword anchor \\
    \textbf{K$_L$} & \textit{keyword}|\textcolor{orange}{[}\textit{keyword}+\textcolor{orange}{]} & keyword literals\\
    \textbf{K$_F$} & \textcolor{orange}{OF} (\textcolor{orange}{(}\textbf{C}\textcolor{orange}{)}|\textcolor{orange}{(}\textbf{J}\textcolor{orange}{)}|\textcolor{orange}{(}\textbf{PU}\textcolor{orange}{)}|\textcolor{orange}{(}\textbf{PE}\textcolor{orange}{)}|\textbf{C$_L$}|\textbf{J$_L$}|\textbf{PU$_L$}|\textbf{PE$_L$})& keyword filters\\\hline
   
    \textbf{PU} & \textit{limit}? (\textbf{PU$_O$}|(\textcolor{orange}{$\sim$}\textit{rank})? (\textcolor{orange}{MOST CITED}|\textcolor{orange}{NEWEST}|\textcolor{orange}{OLDEST}) (\textcolor{orange}{(}\textbf{PU}\textcolor{orange}{)}|\textbf{PU$_L$})) & publication anchor\\
    \textbf{PU$_O$} & (\textcolor{orange}{PUBLICATION}|\textcolor{orange}{BOOK}|\textcolor{orange}{ARTICLE}|\textcolor{orange}{PHDTHESIS}|\textcolor{orange}{MASTERTHESIS}|\textcolor{orange}{INPROCEEDING}|\textcolor{orange}{INCOLLECTION}|\textcolor{orange}{PROCEEDING}) \textbf{PU$_F$}? & publication object\\
    \textbf{PU$_L$} & \textit{dblpKey}|\textit{DOI}|\textcolor{orange}{$\sim$}? \textit{title} & publication literals\\
    \textbf{PU$_F$} & (\textbf{PU$_A$}|\textcolor{orange}{WITH DBLPKEY} \textit{dblpKey}|\textcolor{orange}{WITH DOI} \textit{doi}|\textcolor{orange}{WITH ISBN} \textit{isbn}|\textcolor{orange}{TITLED $\sim$}? \textit{title}|\textcolor{orange}{ABOUT} (\textcolor{orange}{KEYWORD} (\textcolor{orange}{(}\textbf{K}\textcolor{orange}{)}|\textbf{K$_L$}) |\textcolor{orange}{TERMS} \textit{search\_terms}))|\textcolor{orange}{WITH YEAR} \textbf{COMP} \textit{year}|\textcolor{orange}{APPEARED IN} (\textcolor{orange}{(}\textbf{C}\textcolor{orange}{)}|\textbf{C$_L$}|\textcolor{orange}{(}\textbf{J}\textcolor{orange}{)}|\textbf{J$_L$})|(\textcolor{orange}{CITED BY}| \textcolor{orange}{REFERENCES} (\textcolor{orange}{(}\textbf{PU}\textcolor{orange}{)}|\textbf{PU$_L$})|(\textcolor{orange}{EDITED}|\textcolor{orange}{WRITTEN}) \textcolor{orange}{BY} (\textcolor{orange}{(}\textbf{PE}\textcolor{orange}{)}|\textbf{PE$_L$})|\textcolor{orange}{WRITTEN BY ANY DISTINCT}? \textit{number} \textcolor{orange}{OF [}(\textcolor{orange}{(}\textbf{PE}\textcolor{orange}{)}|\textbf{PE$_L$})+\textcolor{orange}{]}|\textcolor{orange}{PUBLISHED WITH} (\textcolor{orange}{(}\textbf{I}\textcolor{orange}{)}|\textbf{I$_L$})))|(\textcolor{orange}{WITH} (((\textcolor{orange}{$\sim$}\textit{rank})? (\textcolor{orange}{MOST}|\textcolor{orange}{LEAST}) ((\textcolor{orange}{REFERENCES} (\textcolor{orange}{TO} (\textcolor{orange}{(}\textbf{PU}\textcolor{orange}{)}|\textbf{PU$_L$}))?)|(\textcolor{orange}{CITATIONS} (\textcolor{orange}{FROM} (\textcolor{orange}{(}\textbf{PU}\textcolor{orange}{)}|\textbf{PU$_L$}))?)))|((\textcolor{orange}{$\sim$}\textit{rank})? ((\textcolor{orange}{HIGHEST}| \textcolor{orange}{LOWEST}) \textbf{METRIC})|(\textcolor{orange}{LONGEST}|\textcolor{orange}{SHORTEST}) (\textcolor{orange}{title}|\textcolor{orange}{abstract}))|(\textbf{METRIC} \textbf{COMP} (\textit{h\_index}|\textit{core\_rank}))|((\textcolor{orange}{title}|\textcolor{orange}{abstract}) \textcolor{orange}{LENGTH} \textbf{COMP} \textit{number})|(\textbf{COMP} \textit{number} ((\textcolor{orange}{REFERENCES} (\textcolor{orange}{TO} (\textcolor{orange}{(}\textbf{PU}\textcolor{orange}{)}|\textbf{PU$_L$}))?)| (\textcolor{orange}{CITATIONS} (\textcolor{orange}{FROM} (\textcolor{orange}{(}\textbf{PU}\textcolor{orange}{)}|\textbf{PU$_L$}))?))))) (\textbf{OPS} \textbf{PU$_F$})? & publication filters\\\hline
    
    \textbf{PE} & \textit{limit}? (\textbf{PE$_O$}|(\textcolor{orange}{COAUTHORS OF} 
    (\textcolor{orange}{(}\textbf{PE}\textcolor{orange}{)}|\textbf{PE$_L$}))|((\textcolor{orange}{$\sim$}\textit{rank})? \textcolor{orange}{MOST} (\textcolor{orange}{PUBLISHING} (\textcolor{orange}{(}\textbf{PE}\textcolor{orange}{)}|\textbf{PE$_L$}) \textcolor{orange}{IN} (\textcolor{orange}{(}\textbf{C}\textcolor{orange}{)}|\textbf{C$_L$}|\textcolor{orange}{(}\textbf{J}\textcolor{orange}{)}|\textbf{J$_L$}))|\textcolor{orange}{RESEARCHING} (\textcolor{orange}{(}\textbf{PE}\textcolor{orange}{)}|\textbf{PE$_L$}) \textcolor{orange}{ABOUT} (\textcolor{orange}{(}\textbf{K}\textcolor{orange}{)}|\textbf{K$_L$}))) & person anchor\\
    \textbf{PE$_O$} & (\textcolor{orange}{PERSON}|\textcolor{orange}{AUTHOR}|\textcolor{orange}{EDITOR}) \textbf{PE$_F$}?& person object\\    
    \textbf{PE$_L$} & \textit{dblpKey}|\textit{ORCID}|(\textcolor{orange}{$\sim$}|\textcolor{orange}{=})? \textit{name} & person literals\\
    \textbf{PE$_F$} & (\textbf{PE$_A$}|\textcolor{orange}{WITH DBLPKEY} \textit{dblpKey}|\textcolor{orange}{NAMED} (\textcolor{orange}{$\sim$}|\textcolor{orange}{=})? \textit{name}|\textcolor{orange}{WITH ORCID} \textit{orcid}|\textcolor{orange}{AUTHORED} (\textcolor{orange}{ONLY}|\textcolor{orange}{NO})? (\textcolor{orange}{(}\textbf{PU}\textcolor{orange}{)}|\textbf{PU$_L$})|\textcolor{orange}{EDITED} (\textcolor{orange}{(}\textbf{PU}\textcolor{orange}{)}|\textbf{PU$_L$})|(\textcolor{orange}{CITED BY}|\textcolor{orange}{REFERENCES}) (\textcolor{orange}{(}\textbf{PU}\textcolor{orange}{)}|\textbf{PU$_L$})|\textcolor{orange}{WORKS FOR} (\textcolor{orange}{(}\textbf{I}\textcolor{orange}{)}|\textbf{I$_L$})|\textcolor{orange}{PUBLISHED IN} (\textcolor{orange}{(}\textbf{C}\textcolor{orange}{)}|\textbf{C$_L$}|\textcolor{orange}{(}\textbf{J}\textcolor{orange}{)}|\textbf{J$_L$}))|(\textcolor{orange}{WITH} (((\textcolor{orange}{$\sim$}\textit{rank})? (\textcolor{orange}{MOST}|\textcolor{orange}{LEAST}) ((\textcolor{orange}{REFERENCES} (\textcolor{orange}{TO} (\textcolor{orange}{(}\textbf{PU}\textcolor{orange}{)}|\textbf{PU$_L$}))?)|(\textcolor{orange}{CITATIONS} (\textcolor{orange}{FROM} (\textcolor{orange}{(}\textbf{PU}\textcolor{orange}{)}|\textbf{PU$_L$}))?)))|((\textcolor{orange}{$\sim$}\textit{rank})? ((\textcolor{orange}{HIGHEST}|\textcolor{orange}{LOWEST}) \textbf{METRIC})|((\textcolor{orange}{LONGEST}|\textcolor{orange}{SHORTEST}) \textcolor{orange}{name})|((\textcolor{orange}{MOST}|\textcolor{orange}{LEAST}) \textcolor{orange}{COAUTHORS} (\textcolor{orange}{IN} (\textcolor{orange}{(}\textbf{PU}\textcolor{orange}{)}|\textbf{PU$_L$}))?))|(\textbf{METRIC} \textbf{COMP} (\textit{h\_index}|\textit{core\_rank}))| (\textbf{COMP} \textit{number} \textcolor{orange}{COAUTHORS} (\textcolor{orange}{IN} (\textcolor{orange}{(}\textbf{PU}\textcolor{orange}{)}|\textbf{PU$_L$}))?)|(\textcolor{orange}{name} \textcolor{orange}{LENGTH} \textbf{COMP} \textit{number}))) (\textbf{OPS} \textbf{PE$_F$})? & person filters\\ \hline

    \textbf{I} & \textit{limit}? (\textbf{I$_O$}|((\textcolor{orange}{$\sim$}\textit{rank})? \textcolor{orange}{MOST RESEARCHING} (\textcolor{orange}{(}\textbf{I}\textcolor{orange}{)}|\textbf{I$_L$}) \textcolor{orange}{ABOUT} (\textcolor{orange}{(}\textbf{K}\textcolor{orange}{)}|\textbf{K$_L$}))) & institution anchor\\
    \textbf{I$_O$} & \textcolor{orange}{INSTITUTION} \textbf{I$_F$}? & institution object\\
    \textbf{I$_L$} & \textit{dblpKey}|\textcolor{orange}{$\sim$}? \textit{name} & institution literals\\
    \textbf{I$_F$} & (\textcolor{orange}{WITH DBLPKEY} \textit{dblpKey}|\textcolor{orange}{NAMED} \textcolor{orange}{$\sim$}? \textit{name}|\textcolor{orange}{WITH CITY} \textit{city}|\textcolor{orange}{WITH COUNTRY} \textit{country}|\textcolor{orange}{WITH MEMBERS} (\textcolor{orange}{(}\textbf{PE}\textcolor{orange}{)}|\textbf{PE$_L$}))|(\textcolor{orange}{WITH} (((\textcolor{orange}{$\sim$}\textit{rank})? ((\textcolor{orange}{HIGHEST}| \textcolor{orange}{LOWEST}) \textbf{METRIC})|((\textcolor{orange}{LONGEST}|\textcolor{orange}{SHORTEST}) (\textcolor{orange}{name}|\textcolor{orange}{location}))| (\textbf{METRIC} \textbf{COMP} (\textit{h\_index}|\textit{core\_rank}))|((\textcolor{orange}{name}|\textcolor{orange}{(}\textcolor{orange}{location}) \textcolor{orange}{LENGTH} \textbf{COMP} \textit{number}))) (\textbf{OPS} \textbf{I$_F$})? & institution filters\\\hline

    \textbf{F} & (\textcolor{orange}{COUNT} \textcolor{orange}{(}\textbf{query}\textcolor{orange}{)})|(\textcolor{orange}{CORE RANKS FOR} (\textcolor{orange}{(}\textbf{PE}\textcolor{orange}{)}|\textbf{PE$_L$}) (\textcolor{orange}{IN} (\textcolor{orange}{(}\textbf{C}\textcolor{orange}{)}|\textcolor{orange}{(}\textbf{J}\textcolor{orange}{)}|\textcolor{orange}{(}\textbf{PU}\textcolor{orange}{)}|\textbf{C$_L$}|\textbf{J$_L$}|\textbf{PU$_L$}))?)|(\textit{limit}? \textcolor{orange}{ALTERNATIVE NAMES FOR} (\textcolor{orange}{(}\textbf{C}\textcolor{orange}{)}|\textcolor{orange}{(}\textbf{J}\textcolor{orange}{)}|\textcolor{orange}{(}\textbf{I}\textcolor{orange}{)}|\textcolor{orange}{(}\textbf{PE}\textcolor{orange}{)}|\textbf{C$_L$}|\textbf{J$_L$}|\textbf{I$_L$}|\textbf{PE$_L$})|(\textit{limit}? \textcolor{orange}{MOST FREQUENT} \textit{attribute} \textcolor{orange}{OF (}\textbf{query}\textcolor{orange}{)})|(\textit{limit}? \textcolor{orange}{[}\textit{attribute}+\textcolor{orange}{]} \textcolor{orange}{OF (}\textbf{query}\textcolor{orange}{)}))& general functions\\ \hline
    \textbf{METRIC} & (\textcolor{orange}{CORERANK METRIC})|(\textcolor{orange}{H-AVG METRIC})& metric\\
   
    \textbf{COMP} & (\textcolor{orange}{AT LEAST}|\textcolor{orange}{AT MOST}|\textcolor{orange}{MORE THAN}|\textcolor{orange}{LESS THAN})?  & comparator\\
    \textbf{OPS} & \textcolor{orange}{AND}|\textcolor{orange}{OR}|\textcolor{orange}{AND NOT}|\textcolor{orange}{OR NOT} & operators\\
\end{tabular}
\caption{SchenQL grammar. Bold terms represent starting points for new rules, orange terms represent fixed language components, italic terms are variables of the type they describe. Black brackets, pipe symbols, plus symbols and question marks are only used for readability of the grammar. Brackets indicate scopes which signs refer to. Pipes indicate a choice between language components. Plus symbols indicate a comma-separated list of the preceding component. Question marks indicate that the preceding language component is optional.}
\label{tab:grammar}
\end{table*}

\section{SchenQL}
SchenQL~\cite{schenqljournal} describes both the underlying domain-specific query language and the graphical user interface. We extend its previously presented functional range. Our main goal is to support all types of users of digital libraries in their information needs, even domain-experts who tend to use more sophisticated search options compared to casual users~\cite{zavalina}. Thus, we offer a broad selection of domain-specific functionalities.
Our system supports query formulation as well as information exploration but focuses more on the formulation as users of digital libraries have been found to prefer searching to browsing in some domains~\cite{zavalina}.
While our query language can be applied to any digital library, it has been designed with the dblp computer science bibliography\footnote{\url{https://dblp.org/}} as central use case. Therefore, some language components model the particularities of dblp, such as using keys for referencing publications and persons (e.g., ``homepages/f/EdwardAFox'') and using numerical suffixes for disambiguating author names (e.g., ``Wei Wang 0042''). 

\subsection{Grammar}


We extend and redefine 
SchenQL~\cite{schenqljournal}, but all previously described functionality remains (a hidden) part of it. On top of that, we add significant extensions that comprise more and more powerful aggregations, more and more powerful functions (that were previously called filters), and a modified syntax that resembles natural language even more.

Table~\ref{tab:grammar} presents the grammar of SchenQL in a compact form. SchenQL consists of the following 6 \textbf{base concepts} which are interrelated: conferences, journals, keywords, publications, persons and institutions. Queries can be constructed from these base concepts. Here, singular and plural of base concepts are possible (e.g. \texttt{PERSON} and \texttt{PERSONS}). If a base concept is used, entities of this type are retrieved for that part of a query. Occurrences of base concepts in queries can be replaced by \textbf{specialisations} of these base concepts (e.g. \texttt{BOOK} instead of \texttt{PUBLICATION}). 

\textbf{Filters} on base concepts restrict the retrieved entities of the target base concept type (e.g. \texttt{CONFERENCE WITH ACRONYM "JCDL"} restricts the result to conferences with the specified acronym or \texttt{ARTICLE WITH YEAR 2022} restricts the result to publications of type article which appeared in 2022). Filters start with a \texttt{WITH} to help resemble natural language if the respective function describes an attribute that entities of the base concept \textit{have}. For example, a publication \textit{has} a year (thus \texttt{PUBLICATION WITH YEAR 2022}), but a publication \textit{is} about keywords (thus \texttt{PUBLICATION ABOUT KEYWORD "dsql"}). Some filters are aggregations such as \texttt{ARTICLES WITH HIGHEST CORERANK METRIC} or \texttt{AUTHORS WITH LEAST COAUTHORS}.
When observing keywords or words, we offer full-text search for words over titles and abstract (e.g. \texttt{PUBLICATIONS ABOUT TERMS "(digital:libraries)|dsql"}), while keywords are a list of topics associated with the publication (e.g. \texttt{PUBLICATIONS ABOUT KEY\-WORDS ["digital libraries", "dsql"]}).

Filters on base concepts can be concatenated with Boolean \textbf{operators} (e.g. \texttt{CONFERENCE WITH ACRONYM "JCDL" OR WITH YEAR 2000} or \texttt{INSTITUTIONS WITH COUNTRY "DE" AND NOT WITH CITY "Cologne"}). An \texttt{AND} operator is implicitly assumed between multiple filters and can be dropped (e.g. \texttt{ARTICLES WITH YEAR AT LEAST 2010 AND WITH YEAR AT MOST 2020} is equivalent to \texttt{ARTICLES WITH YEAR AT LEAST 2010 WITH YEAR AT MOST 2020}).

\textbf{Literals} can be utilised to identify single or multiple entities of a specific base type. They can be used in filters instead of the longer formulation mentioning the base concept specifically (e.g. shortening \texttt{PUBLICATIONS APPEARED IN (CONFERENCE WITH ACRONYM "JCDL")} to \texttt{PUBLICATIONS APPEARED IN "JCDL"} by using the acronym literal).

To enable different types of searches based on string literals like author names, certain prefixes can be used: \texttt{$\sim$} indicates a search, where all blank-separated specified query components need to be part of the name in any order (e.g. \texttt{PERSONS NAMED $\sim$"wang wei"} returns persons with names \textit{Wei-Bo Wang} and \textit{Wang Wei Lee}), while incorporation of \texttt{=} before person names indicates a required exact match of the name. If no prefix is used for search of person names, names with and without the numerical suffix are retrieved 
(e.g. \texttt{PERSONS NAMED ="Wei Wang"} returns the one person with the unique name \textit{Wei Wang} but does not return \textit{Wei Wang 0042} while \texttt{PERSONS NAMED "Wei Wang"} returns both of them). 

\textbf{Restriction} of the number of returned entities of queries can be achieved by either using \texttt{LIMIT} or \texttt{$\sim$RANK} in specific positions in the grammar (see Table~\ref{tab:grammar}). The first variant specifies an upper limit for the number of results to return (e.g. \texttt{5 ARTICLES CITED BY "journals/interactions/Myers98"}), the second variant observes ranks of resulting entities such that more than the specified amount of results can be returned if multiple occupy the same rank (e.g. \texttt{CONFERENCES WITH $\sim$5 HIGHEST H-AVG METRIC}, which may return more than five results if some top conferences have the same H-AVG metric). Here, an application of the rank restriction is only possible if something is aggregated either by using an aggregation filter or the respective anchor points for keywords, publications, persons or institutions (via e.g. \texttt{MOST CITED} or \texttt{MOST RESEARCHING}). 

SchenQL offers a number of \textbf{general functions} on bibliographic metadata that do not return entities of base concepts but other data: \texttt{COUNT} returns the number of entities in the following query, \texttt{CORE RANKS FOR} lists the core ranks and the number of occurrences for the following entities, \texttt{ALTERNATIVE NAMES FOR} returns the list of names for the following entities, \texttt{MOST FREQUENT} \textit{attribute} \texttt{OF} returns the most frequent value of the specified attribute in the following query and \textit{attribute} \texttt{OF} returns the values for the attributes of the following query.

The complete overview of SchenQL's grammar with filters and query components can be found in Table~\ref{tab:grammar}. Values of non-numeric variables always need to be encompassed by quotation marks.

\subsection{GUI}

\begin{figure}
\centering
    \includegraphics[width=0.47\textwidth]{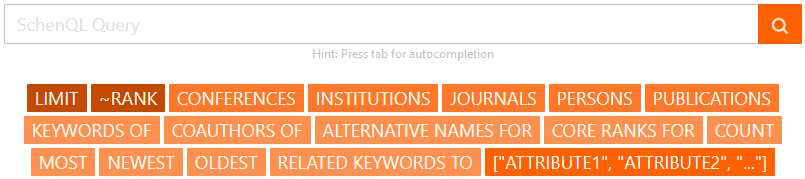}
    \caption{SchenQL search interface with colour-coded (dependent on the component type) query component suggestions.}
    \label{fig:search}
\end{figure}

\begin{figure}
    \centering
    \includegraphics[width=0.47\textwidth]{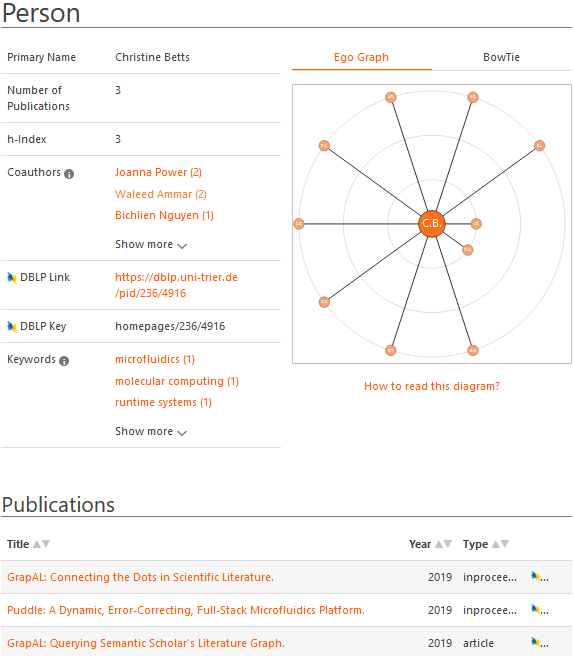}
    \caption{Person detail page with Ego Graph and BowTie visualisations.}
    \label{fig:person}
\end{figure}

\begin{figure}
    \centering
    \includegraphics[width=0.47\textwidth]{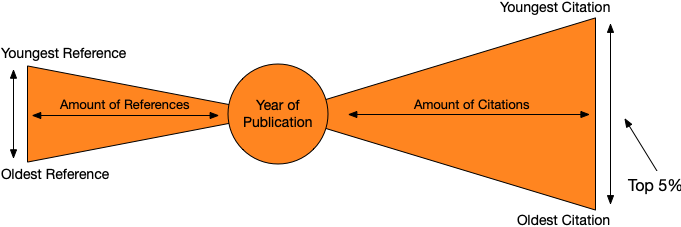}
    \caption{BowTie visualisation~\cite{schenqljournal} explained.}

    \label{fig:bowtie}
\end{figure}

The SchenQL GUI is an updated form of the SchenQL GUI~\cite{schenqljournal}. The search interface supports query construction with a suggestion and auto-completion feature (shown in Figure~\ref{fig:search}). Possible components to continue queries at any given point are suggested below, they can be clicked to appear as part of queries. In the component suggestions, we did not include the specialisations as this would overload the interface. Possible following language components are colour-coded. The colour indicates the type of the component e.g. base concepts (such as \texttt{CONFERENCES} or \texttt{PERSONS}), filters (such as \texttt{WITH YEAR} or \texttt{ABOUT}), literals (such as \texttt{NUMBER} or \texttt{"STRING"}), or restrictions (such as \texttt{LIMIT} or \texttt{$\sim$RANK}).
After running a SchenQL query, a search result is shown. Clicking on result items takes a user to the detail view of the specific type, Figure~\ref{fig:person} shows a detail view of a person. Here we include details on the publications, citation information and keywords of their work. Additionally, the Ego Graph~\cite{DBLP:journals/corr/abs-1009-5183} visualisation (in the right part of Figure~\ref{fig:person}) highlights the most frequent co-authors of a person in the middle. The nodes surrounding the middle symbolise the co-authors, the shorter the distance between the middle and the co-author, the more works have been published together. Another visualisation included in person's, publication's, conference's and journal's detail views is the BowTie~\cite{DBLP:journals/jodl/KhazaeiH17}. It can be used to estimate an entity's position in the scientific landscape, as it visualises the number and age distribution of references and citations (see Figure~\ref{fig:bowtie} for a visual explanation).

\begin{table*}
\scriptsize
    \centering
    \begin{tabular}{l|p{11cm}}
\textbf{Information need} & \textbf{SchenQL query formulation}\\ \hline
     Find author by name~\cite{DBLP:conf/acl/BettsPA19} & PERSON NAMED "Christine Betts"\\
     Fuzzy author name search~\cite{DBLP:conf/acl/BettsPA19} & PERSON NAMED $\sim$"Wei Wang"\\
     Find authors of paper~\cite{DBLP:conf/acl/BettsPA19,DBLP:conf/ercimdl/BloehdornCDHHTV07,zhu} & PERSONS AUTHORED "conf/acl/BettsPA19"\\
     Find most active author on topic in time frame~\cite{DBLP:conf/acl/BettsPA19} & MOST RESEARCHING (PERSON AUTHORED (PUBLICATIONS WITH YEAR AT LEAST 2000 WITH YEAR AT MOST 2020)) ABOUT KEYWORDS "digital libraries"\\ 
     Find co-authors~\cite{zhu,DBLP:conf/edbt/Gomez-VillamorSGMML08} & COAUTHORS OF "Adam Jatowt"\\

     Find articles of author in venue~\cite{DBLP:conf/ercimdl/BloehdornCDHHTV07} & ARTICLES WRITTEN BY "Waleed Ammar" APPEARED IN "JCDL"\\
     Find papers citing author~\cite{zhu} & PUBLICATIONS REFERENCES (PUBLICATIONS WRITTEN BY "Yongjun Zhu" )\\
     Find papers on topics~\cite{DBLP:conf/acl/BettsPA19,DBLP:conf/ercimdl/BloehdornCDHHTV07,zhu} & PUBLICATIONS ABOUT KEYWORD ["digital libraries", "search"]\\
     Find papers using keywords~\cite{DBLP:conf/acl/BettsPA19} & PUBLICATIONS ABOUT TERMS "digital:libraries|dsql"\\
     Find papers on topic cited by author's papers in venue~\cite{zhu} & PUBLICATIONS ABOUT KEYWORD "search" CITED BY (PUBLICATIONS WRITTEN BY "Joeran Beel" APPEARED IN "JCDL")\\
     Find number of papers on topic \textit{a} referencing papers on topic \textit{b}~\cite{DBLP:conf/acl/BettsPA19} & COUNT (PUBLICATION ABOUT KEYWORD "digital library" REFERENCES (PUBLICATIONS ABOUT KEYWORD "search")\\

     
     Find institution of authors of paper~\cite{zhu} & INSTITUTIONS WITH MEMBERS (PERSONS AUTHORED "conf/cikm/ZhuSRH08")\\
     
     Find topics of author's papers in venue~\cite{zhu} & KEYWORDS OF (PUBLICATIONS APPEARED IN "JCDL" WRITTEN BY "Michael Ley")\\
     Find topics of papers cited by papers on topic \textit{a}~\cite{zhu} & KEYWORDS OF (PUBLICATIONS CITED BY (PUBLICATIONS ABOUT KEYWORD "dsql"))\\ 
     
    \hline

     Find persons with 5 longest names & PERSONS WITH $\sim$5 LONGEST NAME\\
     Find authors of 5 earliest papers on topic & PERSONS AUTHORED $(\sim$5 OLDEST (PUBLICATIONS ABOUT KEYWORD "digital library"))\\
     Find authors who only recently started working on topic & PERSONS AUTHORED (PUBLICATIONS ABOUT "digital libraries") AND AUTHORED NO (PUBLICATIONS ABOUT KEYWORD "digital libraries" WITH YEAR AT MOST 2017) \\
     Find authors who published the most on topic since year~\cite{DBLP:conf/acl/BettsPA19} & MOST RESEARCHING (PERSONS AUTHORED (PUBLICATIONS WITH YEAR AT LEAST 2015)) ABOUT KEYWORD "dsql"\\
     Find author co-citation~\cite{zhu} & PUBLICATIONS REFERENCES (PUBLICATIONS WRITTEN BY "Ralf Schenkel") REFERENCES (PUBLICATIONS WRITTEN BY "Norbert Fuhr")\\

     Find papers written by members of institutions \textit{a} and \textit{b} & PUBLICATIONS WRITTEN BY ANY DISTINCT 2 OF [(PERSON WORKS FOR "University of Pisa"), (PERSON WORKS FOR "National Research Council, Italy")] \\
     Find papers from year cited at least 20 times & PUBLICATIONS WITH YEAR 2020 WITH AT LEAST 20 CITATIONS\\
     Find papers close to area~\cite{DBLP:conf/edbt/Gomez-VillamorSGMML08} & PUBLICATIONS ABOUT ($\sim$ 5 MOST FREQUENT KEYWORDS OF (RELATED KEYWORDS TO "digital libraries"))\\
    
     Find institution in country with highest h index & INSTITUTIONS WITH COUNTRY "DE" WITH HIGHEST H-AVG METRIC\\

     Find topics of papers in conference in year & KEYWORDS OF (PUBLICATIONS WITH YEAR 2018 APPEARED IN "JCDL") \\
     Find co-words~\cite{zhu} & RELATED KEYWORDS TO "digital libraries" \\

     Find most cited paper in journal & MOST CITED (PUBLICATION APPEARED IN "JODL")\\
     Find bibliographic couplings~\cite{zhu} & PUBLICATIONS CITED BY "conf/aaai/ChangRRR08" CITED BY "conf/aaai/HashemiH18"\\
     Find co-cited papers~\cite{zhu} & PUBLICATIONS REFERENCES "journals/annals/Grudin05" REFERENCES "journals/interactions/Myers98"\\

    \end{tabular}
    \caption{Common (upper part) and specialised (lower part) information needs of users of digital libraries and SchenQL queries.}
    \label{tab:queries}
\end{table*}

\subsection{System Architecture and Data}

\begin{figure}
    \centering
    \includegraphics[width=0.47\textwidth]{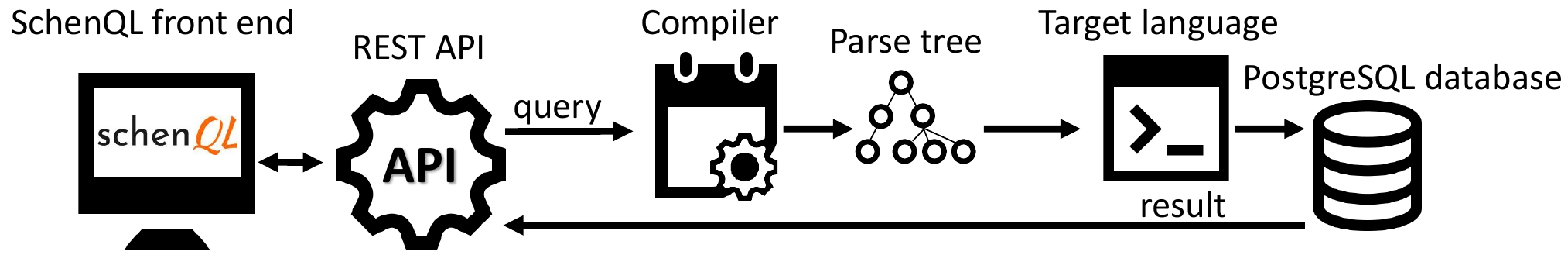}
    \caption{Transformation of SchenQL query to result.}
    \label{schenQLquery}
\end{figure}

The SchenQL front-end uses Node.js 16.13 and incorporates Boot and Spring Boot. It communicates with a REST API which accepts user input queries and relays them to the compiler module. Here, for the lexer and parser, we extended and modified the previous SchenQL grammar which was generated with ANTLR to also support aggregation functions as well as more filters in SchenQL. The SchenQL queries are compiled via Java 17 into an intermediate parse tree structure, before they are translated to an exchangeable target language. We currently run queries against a PostgreSQL\footnote{Compared to the previous version of SchenQL~\cite{schenqljournal}, here we decided to use PostgreSQL instead of MySQL for increased performance.} 14.1 database. Results of queries are transmitted back through the API to the front-end where they are used to populate the result tables and detail views of the GUI (see Figure~\ref{schenQLquery}).

As data source for our system, we combine the dblp XML dump from 1st October 2021\footnote{\url{https://dblp.org/xml/release/dblp-2021-10-01.xml.gz}} with Semantic Scholar data~\cite{DBLP:conf/naacl/AmmarGBBCDDEFHK18} from October 2021 for citation information as well as AMiner Open Academic Graph 2.1~\cite{DBLP:conf/www/SinhaSSMEHW15,DBLP:conf/kdd/TangZYLZS08,DBLP:conf/kdd/ZhangLTDYZGWSLW19} data for identifying automatically generated keywords of publications; abstracts are taken from both collections.

\subsection{Application and Information Needs}

SchenQL can be used to satisfy typical information needs of users of digital libraries or more complex ones which cannot directly be answered by current digital libraries.

Table~\ref{tab:queries} contains a selection of typical information needs (upper part) and ones which cannot be directly satisfied by using digital libraries (lower part) as well as their exemplary formulations in SchenQL. The listed information needs are partially mentioned in previous works~\cite{zhu,DBLP:conf/acl/BettsPA19,DBLP:conf/ercimdl/BloehdornCDHHTV07,DBLP:conf/edbt/Gomez-VillamorSGMML08}.
Further information needs such as query refinement, backwards and forwards citation analysis, quality judgement~\cite{DBLP:journals/oir/HoeberK15} as well as support in query formulation and general information exploration are mainly supported by the GUI of SchenQL.

\section{Conclusion}

We presented an extension of the SchenQL query language and GUI that incorporates aggregation functions and supports more general functions. Our system aims to support casual users as well as domain-experts by providing a huge selection of domain-specific functionalities. We make these functions accessible especially for casual users by suggesting possible following query components in our GUI.

SchenQL could be extended by enabling the support of graph-based operations (as seen with GrapAL~\cite{DBLP:conf/acl/BettsPA19}) such as the calculation of centrality scores, PageRank or hubs and authorities on a user-defined portion of the bibliographic network. Another line of work could focus on the incorporation of different visualisations to support users' needs, for example colour-coded topical distributions in publications to help find relevant papers or estimate an author's fit as a reviewer for a manuscript.

Evaluations of SchenQL could include information search where users are asked to retrieve information for predefined queries such as the ones we presented in Table~\ref{tab:queries} or more open, task-based studies where user behaviour and preferences could be observed more in detail as seen in other evaluations~\cite{DBLP:journals/oir/HoeberK15}. For example sense-making studies where users are asked to find and collect information~\cite{DBLP:journals/computer/Pirolli09} such as the question which one of two researchers could better fit a fictional open academic position, or exploratory search which incorporate information lookup, learning and investigation~\cite{DBLP:journals/computer/Pirolli09} such as asking the users to get familiar with new subject areas could provide interesting insights and highlight further directions.

\bibliographystyle{ACM-Reference-Format}
\bibliography{bib}


\begin{thebibliography}{17}


\ifx \showCODEN    \undefined \def \showCODEN     #1{\unskip}     \fi
\ifx \showDOI      \undefined \def \showDOI       #1{#1}\fi
\ifx \showISBNx    \undefined \def \showISBNx     #1{\unskip}     \fi
\ifx \showISBNxiii \undefined \def \showISBNxiii  #1{\unskip}     \fi
\ifx \showISSN     \undefined \def \showISSN      #1{\unskip}     \fi
\ifx \showLCCN     \undefined \def \showLCCN      #1{\unskip}     \fi
\ifx \shownote     \undefined \def \shownote      #1{#1}          \fi
\ifx \showarticletitle \undefined \def \showarticletitle #1{#1}   \fi
\ifx \showURL      \undefined \def \showURL       {\relax}        \fi
\providecommand\bibfield[2]{#2}
\providecommand\bibinfo[2]{#2}
\providecommand\natexlab[1]{#1}
\providecommand\showeprint[2][]{arXiv:#2}

\bibitem[Ammar et~al\mbox{.}(2018)]%
        {DBLP:conf/naacl/AmmarGBBCDDEFHK18}
\bibfield{author}{\bibinfo{person}{Waleed Ammar}, \bibinfo{person}{Dirk
  Groeneveld}, \bibinfo{person}{Chandra Bhagavatula}, \bibinfo{person}{Iz
  Beltagy}, \bibinfo{person}{Miles Crawford}, \bibinfo{person}{Doug Downey},
  \bibinfo{person}{Jason Dunkelberger}, \bibinfo{person}{Ahmed Elgohary},
  \bibinfo{person}{Sergey Feldman}, \bibinfo{person}{Vu Ha},
  \bibinfo{person}{Rodney Kinney}, \bibinfo{person}{Sebastian Kohlmeier},
  \bibinfo{person}{Kyle Lo}, \bibinfo{person}{Tyler Murray},
  \bibinfo{person}{Hsu{-}Han Ooi}, \bibinfo{person}{Matthew~E. Peters},
  \bibinfo{person}{Joanna Power}, \bibinfo{person}{Sam Skjonsberg},
  \bibinfo{person}{Lucy~Lu Wang}, \bibinfo{person}{Chris Wilhelm},
  \bibinfo{person}{Zheng Yuan}, \bibinfo{person}{Madeleine van Zuylen}, {and}
  \bibinfo{person}{Oren Etzioni}.} \bibinfo{year}{2018}\natexlab{}.
\newblock \showarticletitle{Construction of the Literature Graph in Semantic
  Scholar}. In \bibinfo{booktitle}{\emph{{NAACL-HLT} {(3)}}}.
  \bibinfo{publisher}{Association for Computational Linguistics},
  \bibinfo{pages}{84--91}.
\newblock


\bibitem[Betts et~al\mbox{.}(2019)]%
        {DBLP:conf/acl/BettsPA19}
\bibfield{author}{\bibinfo{person}{Christine Betts}, \bibinfo{person}{Joanna
  Power}, {and} \bibinfo{person}{Waleed Ammar}.}
  \bibinfo{year}{2019}\natexlab{}.
\newblock \showarticletitle{GrapAL: Connecting the Dots in Scientific
  Literature}. In \bibinfo{booktitle}{\emph{Proceedings of the 57th Conference
  of the Association for Computational Linguistics, {ACL} 2019, Florence,
  Italy, July 28 - August 2, 2019, Volume 3: System Demonstrations}},
  \bibfield{editor}{\bibinfo{person}{Marta~R. Costa{-}juss{\`{a}}} {and}
  \bibinfo{person}{Enrique Alfonseca}} (Eds.). \bibinfo{publisher}{Association
  for Computational Linguistics}, \bibinfo{pages}{147--152}.
\newblock
\urldef\tempurl%
\url{https://doi.org/10.18653/v1/p19-3025}
\showDOI{\tempurl}


\bibitem[Bloehdorn et~al\mbox{.}(2007)]%
        {DBLP:conf/ercimdl/BloehdornCDHHTV07}
\bibfield{author}{\bibinfo{person}{Stephan Bloehdorn}, \bibinfo{person}{Philipp
  Cimiano}, \bibinfo{person}{Alistair Duke}, \bibinfo{person}{Peter Haase},
  \bibinfo{person}{J{\"{o}}rg Heizmann}, \bibinfo{person}{Ian Thurlow}, {and}
  \bibinfo{person}{Johanna V{\"{o}}lker}.} \bibinfo{year}{2007}\natexlab{}.
\newblock \showarticletitle{Ontology-Based Question Answering for Digital
  Libraries}. In \bibinfo{booktitle}{\emph{Research and Advanced Technology for
  Digital Libraries, 11th European Conference, {ECDL} 2007, Budapest, Hungary,
  September 16-21, 2007, Proceedings}} \emph{(\bibinfo{series}{Lecture Notes in
  Computer Science}, Vol.~\bibinfo{volume}{4675})},
  \bibfield{editor}{\bibinfo{person}{L{\'{a}}szl{\'{o}} Kov{\'{a}}cs},
  \bibinfo{person}{Norbert Fuhr}, {and} \bibinfo{person}{Carlo Meghini}}
  (Eds.). \bibinfo{publisher}{Springer}, \bibinfo{pages}{14--25}.
\newblock
\urldef\tempurl%
\url{https://doi.org/10.1007/978-3-540-74851-9\_2}
\showDOI{\tempurl}


\bibitem[Francis et~al\mbox{.}(2018)]%
        {cypher}
\bibfield{author}{\bibinfo{person}{Nadime Francis}, \bibinfo{person}{Alastair
  Green}, \bibinfo{person}{Paolo Guagliardo}, \bibinfo{person}{Leonid Libkin},
  \bibinfo{person}{Tobias Lindaaker}, \bibinfo{person}{Victor Marsault},
  \bibinfo{person}{Stefan Plantikow}, \bibinfo{person}{Mats Rydberg},
  \bibinfo{person}{Petra Selmer}, {and} \bibinfo{person}{Andr{\'{e}}s Taylor}.}
  \bibinfo{year}{2018}\natexlab{}.
\newblock \showarticletitle{Cypher: An Evolving Query Language for Property
  Graphs}. In \bibinfo{booktitle}{\emph{Proceedings of the 2018 International
  Conference on Management of Data, {SIGMOD} Conference 2018, Houston, TX, USA,
  June 10-15, 2018}}, \bibfield{editor}{\bibinfo{person}{Gautam Das},
  \bibinfo{person}{Christopher~M. Jermaine}, {and} \bibinfo{person}{Philip~A.
  Bernstein}} (Eds.). \bibinfo{publisher}{{ACM}}, \bibinfo{pages}{1433--1445}.
\newblock
\urldef\tempurl%
\url{https://doi.org/10.1145/3183713.3190657}
\showDOI{\tempurl}


\bibitem[G{\'{o}}mez{-}Villamor et~al\mbox{.}(2008)]%
        {DBLP:conf/edbt/Gomez-VillamorSGMML08}
\bibfield{author}{\bibinfo{person}{Sergio G{\'{o}}mez{-}Villamor},
  \bibinfo{person}{Gerard Soldevila{-}Miranda}, \bibinfo{person}{Aleix
  Gim{\'{e}}nez{-}Va{\~{n}}{\'{o}}}, \bibinfo{person}{Norbert
  Mart{\'{\i}}nez{-}Bazan}, \bibinfo{person}{Victor Munt{\'{e}}s{-}Mulero},
  {and} \bibinfo{person}{Josep~Llu{\'{\i}}s Larriba{-}Pey}.}
  \bibinfo{year}{2008}\natexlab{}.
\newblock \showarticletitle{{BIBEX:} a bibliographic exploration tool based on
  the {DEX} graph query engine}. In \bibinfo{booktitle}{\emph{{EDBT} 2008, 11th
  International Conference on Extending Database Technology, Nantes, France,
  March 25-29, 2008, Proceedings}} \emph{(\bibinfo{series}{{ACM} International
  Conference Proceeding Series}, Vol.~\bibinfo{volume}{261})},
  \bibfield{editor}{\bibinfo{person}{Alfons Kemper}, \bibinfo{person}{Patrick
  Valduriez}, \bibinfo{person}{Noureddine Mouaddib}, \bibinfo{person}{Jens
  Teubner}, \bibinfo{person}{Mokrane Bouzeghoub}, \bibinfo{person}{Volker
  Markl}, \bibinfo{person}{Laurent Amsaleg}, {and} \bibinfo{person}{Ioana
  Manolescu}} (Eds.). \bibinfo{publisher}{{ACM}}, \bibinfo{pages}{735--739}.
\newblock
\urldef\tempurl%
\url{https://doi.org/10.1145/1353343.1353439}
\showDOI{\tempurl}


\bibitem[Hoeber and Khazaei(2015)]%
        {DBLP:journals/oir/HoeberK15}
\bibfield{author}{\bibinfo{person}{Orland Hoeber} {and}
  \bibinfo{person}{Taraneh Khazaei}.} \bibinfo{year}{2015}\natexlab{}.
\newblock \showarticletitle{Evaluating citation visualization and exploration
  methods for supporting academic search tasks}.
\newblock \bibinfo{journal}{\emph{Online Inf. Rev.}} \bibinfo{volume}{39},
  \bibinfo{number}{2} (\bibinfo{year}{2015}), \bibinfo{pages}{229--254}.
\newblock
\urldef\tempurl%
\url{https://doi.org/10.1108/OIR-10-2014-0259}
\showDOI{\tempurl}


\bibitem[Hotho et~al\mbox{.}(2009)]%
        {DBLP:series/xmedia/HothoJBGKSS09}
\bibfield{author}{\bibinfo{person}{Andreas Hotho}, \bibinfo{person}{Robert
  J{\"{a}}schke}, \bibinfo{person}{Dominik Benz}, \bibinfo{person}{Miranda
  Grahl}, \bibinfo{person}{Beate Krause}, \bibinfo{person}{Christoph Schmitz},
  {and} \bibinfo{person}{Gerd Stumme}.} \bibinfo{year}{2009}\natexlab{}.
\newblock \showarticletitle{Social Bookmarking am Beispiel BibSonomy}.
\newblock In \bibinfo{booktitle}{\emph{Social Semantic Web: Web 2.0 - Was
  nun?}}, \bibfield{editor}{\bibinfo{person}{Andreas Blumauer} {and}
  \bibinfo{person}{Tassilo Pellegrini}} (Eds.). \bibinfo{publisher}{Springer},
  \bibinfo{pages}{363--391}.
\newblock
\urldef\tempurl%
\url{https://doi.org/10.1007/978-3-540-72216-8\_18}
\showDOI{\tempurl}


\bibitem[Khazaei and Hoeber(2017)]%
        {DBLP:journals/jodl/KhazaeiH17}
\bibfield{author}{\bibinfo{person}{Taraneh Khazaei} {and}
  \bibinfo{person}{Orland Hoeber}.} \bibinfo{year}{2017}\natexlab{}.
\newblock \showarticletitle{Supporting academic search tasks through citation
  visualization and exploration}.
\newblock \bibinfo{journal}{\emph{Int. J. Digit. Libr.}} \bibinfo{volume}{18},
  \bibinfo{number}{1} (\bibinfo{year}{2017}), \bibinfo{pages}{59--72}.
\newblock
\urldef\tempurl%
\url{https://doi.org/10.1007/s00799-016-0170-x}
\showDOI{\tempurl}


\bibitem[Kreutz et~al\mbox{.}(2021)]%
        {schenqljournal}
\bibfield{author}{\bibinfo{person}{Christin~Katharina Kreutz},
  \bibinfo{person}{Michael Wolz}, \bibinfo{person}{Jascha Knack},
  \bibinfo{person}{Benjamin Weyers}, {and} \bibinfo{person}{Ralf Schenkel}.}
  \bibinfo{year}{2021}\natexlab{}.
\newblock \showarticletitle{SchenQL: in-depth analysis of a query language for
  bibliographic metadata}.
\newblock \bibinfo{journal}{\emph{Int J Digit Libr}} (\bibinfo{year}{2021}).
\newblock
\urldef\tempurl%
\url{https://doi.org/10.1007/s00799-021-00317-8}
\showDOI{\tempurl}


\bibitem[Ley(2009)]%
        {DBLP:journals/pvldb/Ley09}
\bibfield{author}{\bibinfo{person}{Michael Ley}.}
  \bibinfo{year}{2009}\natexlab{}.
\newblock \showarticletitle{{DBLP} - Some Lessons Learned}.
\newblock \bibinfo{journal}{\emph{Proc. {VLDB} Endow.}} \bibinfo{volume}{2},
  \bibinfo{number}{2} (\bibinfo{year}{2009}), \bibinfo{pages}{1493--1500}.
\newblock
\urldef\tempurl%
\url{https://doi.org/10.14778/1687553.1687577}
\showDOI{\tempurl}


\bibitem[Pirolli(2009)]%
        {DBLP:journals/computer/Pirolli09}
\bibfield{author}{\bibinfo{person}{Peter Pirolli}.}
  \bibinfo{year}{2009}\natexlab{}.
\newblock \showarticletitle{Powers of 10: Modeling Complex Information-Seeking
  Systems at Multiple Scales}.
\newblock \bibinfo{journal}{\emph{Computer}} \bibinfo{volume}{42},
  \bibinfo{number}{3} (\bibinfo{year}{2009}), \bibinfo{pages}{33--40}.
\newblock
\urldef\tempurl%
\url{https://doi.org/10.1109/MC.2009.94}
\showDOI{\tempurl}


\bibitem[Reitz(2010)]%
        {DBLP:journals/corr/abs-1009-5183}
\bibfield{author}{\bibinfo{person}{Florian Reitz}.}
  \bibinfo{year}{2010}\natexlab{}.
\newblock \showarticletitle{A Framework for an Ego-centered and Time-aware
  Visualization of Relations in Arbitrary Data Repositories}.
\newblock \bibinfo{journal}{\emph{CoRR}}  \bibinfo{volume}{abs/1009.5183}
  (\bibinfo{year}{2010}).
\newblock
\showeprint[arXiv]{1009.5183}
\urldef\tempurl%
\url{http://arxiv.org/abs/1009.5183}
\showURL{%
\tempurl}


\bibitem[Sinha et~al\mbox{.}(2015)]%
        {DBLP:conf/www/SinhaSSMEHW15}
\bibfield{author}{\bibinfo{person}{Arnab Sinha}, \bibinfo{person}{Zhihong
  Shen}, \bibinfo{person}{Yang Song}, \bibinfo{person}{Hao Ma},
  \bibinfo{person}{Darrin Eide}, \bibinfo{person}{Bo{-}June~Paul Hsu}, {and}
  \bibinfo{person}{Kuansan Wang}.} \bibinfo{year}{2015}\natexlab{}.
\newblock \showarticletitle{An Overview of Microsoft Academic Service {(MAS)}
  and Applications}. In \bibinfo{booktitle}{\emph{{WWW} (Companion Volume)}}.
  \bibinfo{publisher}{{ACM}}, \bibinfo{pages}{243--246}.
\newblock


\bibitem[Tang et~al\mbox{.}(2008)]%
        {DBLP:conf/kdd/TangZYLZS08}
\bibfield{author}{\bibinfo{person}{Jie Tang}, \bibinfo{person}{Jing Zhang},
  \bibinfo{person}{Limin Yao}, \bibinfo{person}{Juanzi Li}, \bibinfo{person}{Li
  Zhang}, {and} \bibinfo{person}{Zhong Su}.} \bibinfo{year}{2008}\natexlab{}.
\newblock \showarticletitle{ArnetMiner: extraction and mining of academic
  social networks}. In \bibinfo{booktitle}{\emph{{KDD}}}.
  \bibinfo{publisher}{{ACM}}, \bibinfo{pages}{990--998}.
\newblock


\bibitem[Zavalina and Vassilieva(2014)]%
        {zavalina}
\bibfield{author}{\bibinfo{person}{Oksana Zavalina} {and}
  \bibinfo{person}{Elena Vassilieva}.} \bibinfo{year}{2014}\natexlab{}.
\newblock \showarticletitle{Understanding the Information Needs of Large-Scale
  Digital Library Users}.
\newblock \bibinfo{journal}{\emph{Library Resources \& Technical Services}}
  \bibinfo{volume}{58} (\bibinfo{date}{04} \bibinfo{year}{2014}),
  \bibinfo{pages}{84}.
\newblock
\urldef\tempurl%
\url{https://doi.org/10.5860/lrts.58n2.84}
\showDOI{\tempurl}


\bibitem[Zhang et~al\mbox{.}(2019)]%
        {DBLP:conf/kdd/ZhangLTDYZGWSLW19}
\bibfield{author}{\bibinfo{person}{Fanjin Zhang}, \bibinfo{person}{Xiao Liu},
  \bibinfo{person}{Jie Tang}, \bibinfo{person}{Yuxiao Dong},
  \bibinfo{person}{Peiran Yao}, \bibinfo{person}{Jie Zhang},
  \bibinfo{person}{Xiaotao Gu}, \bibinfo{person}{Yan Wang},
  \bibinfo{person}{Bin Shao}, \bibinfo{person}{Rui Li}, {and}
  \bibinfo{person}{Kuansan Wang}.} \bibinfo{year}{2019}\natexlab{}.
\newblock \showarticletitle{{OAG:} Toward Linking Large-scale Heterogeneous
  Entity Graphs}. In \bibinfo{booktitle}{\emph{{KDD}}}.
  \bibinfo{publisher}{{ACM}}, \bibinfo{pages}{2585--2595}.
\newblock


\bibitem[Zhu(2017)]%
        {zhu}
\bibfield{author}{\bibinfo{person}{Yongjun Zhu}.}
  \bibinfo{year}{2017}\natexlab{}.
\newblock \bibinfo{title}{Graph-based Interactive Bibliographic Information
  Retrieval Systems}.
\newblock
\newblock
\showISBNx{978-1-3696-8035-5}


\end{thebibliography}

\end{document}